\documentclass[12pt]{article}
\usepackage{graphicx}
\usepackage{graphpap}
\usepackage{epsf}
\usepackage{amssymb}

\evensidemargin=\oddsidemargin \textheight=23cm \textwidth=17.cm

\def\nifsx{${}^{56}$Ni}
\def\fefsx{${}^{56}$Fe}

\def\FIG #1 #2 [#3] #4\par{%
  \begin{figure}[!h] \begin{center}%
    \includegraphics*[#3]{#2}%
    \\ \parbox{10cm}{%
    \caption{\label{#1}#4}}%
  \end{center}\end{figure}%
}

\def\FIGG #1 #2 #3 [#4] #5\par{%
  \begin{figure}[!h]
    \includegraphics*[#4]{#2}
    \hfill
    \includegraphics*[#4]{#3}
    \caption{\label{#1}#5}
  \end{figure}
}

\def\rfig#1{Fig.\ref{#1}}

\renewcommand{\t}{\,\mbox}

\title{Time-dependent thermal X-ray afterglows from GRBs}
\author{D.I.~Kosenko${}^1$, S.I.~Blinnikov${}^{1,2}$,
K.A.~Postnov${}^1$, \\
P.~Lundqvist${}^3$, E.I.~Sorokina${}^1$ \\}

\begin{document}

\maketitle

%\thispagestyle{empty}
%\vskip 1cm

\begin{center}
{\it
${}^1$ Sternberg Astronomical Institute, 119992, Moscow, Russia \\
${}^2$ Institute for Experimental and Theoretical Physics, 117218,
Moscow, Russia \\
${}^3$ Stockholm Observatory, Albanova, Stockholm, Sweden\\
}
lisett@xray.sai.msu.ru, blinn@sai.msu.su, pk@sai.msu.ru, \\
peter@astro.su.se, sorokina@sai.msu.su \\
\end{center}

\begin{abstract}
Time-dependent thermal X-ray spectra are calculated from
physically plausible conditions around GRB. It is shown that
account for time-dependent ionization processes strongly affects
the observed spectra of hot rarefied plasma. These calculations
may provide an alternative explanation to the observed X-ray lines
of early GRBs afterglows (such as GRB 011211). Our technique will
allow one to obtain independent constraints on the GRB collimation
angle and on the clumpiness of circumstellar matter.
\end{abstract}

\section{Introduction}

X-ray afterglows of GRBs were first discovered by Beppo-Sax
satellite (Costa et al. 1997) and their spectral observations are
now used for plasma diagnostic of the plasma surrounding GRB cites
(Lazzati et al. 1999; Piro et al. 2000).
The most exciting
observations are now available from {\it Chandra} and {\it
XMM-Newton} X-ray satellites. Recently, Reeves et al. (2002a,b)
reported on {\it XMM-Newton} observations of the early afterglow
spectrum of GRB 011211. These observations, taken $\sim 11$ hours
after the GRB, revealed the presence of emission lines which were
fading over a time scale shorter than $10^{4}$ s. These lines were
interpreted as blue-shifted (with a velocity of $\sim 0.1 c$ in
the GRB comoving frame) $K_\alpha$ emission from ions of $\alpha$-elements
(Mg XI, Si XIV, S XVI, Ar XVIII, Ca XX) arising in a dense
($n_e\sim 10^{15}\t{cm}^{-3}$), optically-thin thermal plasma
shell heated by the GRB. The model requires a factor of 10
overabundance of $\alpha$-metals in the plasma. No iron lines were
seen, however.

If this interpretation is correct (see the criticism in Borozdin
and Trudolyubov 2002; Rutledge and Sako 2002), this could strongly
point to a supernova (SN) explosion preceding the GRB (the
supranova model; Vietri and Stella 1998). The apparent absence of
iron lines is explained by Reeves et al. by possible peculiarities
of the SN ejecta. It is worth noting the claims (Watson et al.
2002) that three of four GRB X-ray afterglows observed by {\it
XMM-Newton} are best-fitted by thermal plasma model.

The emission line origin in GRB 011211 was critically considered
in several papers (Kumar and Narayan 2002, Lazzati 2002). Our
study (which we have reported at this meeting before the latter
paper was posted in astro-ph e-print archive)
re-examines the X-ray emission line origin in thermal
plasma model. In contrast to other studies, we investigate
time-dependent ionization effects in thermal plasma. We show
inconsistency of the model proposed by Reeves et al. (2002a) for
GRB 011211 and propose another interpretation of what was
observed. Specifically, in our model the observed non-stationarity
of emission lines is directly related to time-dependent character
of collisional ionization in thermal optically thin plasma with
clumps of lower density ($n_e\sim 10^{11}-10^{12}\t{cm}^3$). Such a
plasma can be found around  the GRB or just spuriously occur near our
line of sight.

However, in our model it is difficult to explain the observed blueshift
of lines  by a supernova preceding the GRB for only a few days.
In the thermal model,  no individual lines can be reproduced from fresh
hydrogen-helium SN ejecta even enriched in metals
(there is no space near the GRB source to get the needed emission measure).
Purely metal ejecta are much more efficient X-ray emitters, they can be placed
closer to the GRB, but then, as our computations show,
one would rather observe the metal photorecombination edges, not
lines,  from a recent (a few days old) SN ejecta.
The ten-times metal overabundance is
also unnatural for fresh ejecta, because modern paradigm of massive star SN explosions
assumes purely metal ejecta (for SNe~Ic).
If real, the blueshifted lines should be emitted at much larger distance from
the GRB engine, so they could originate in the material expelled by a supernova
much earlier, maybe some years before the GRB, and this material can be mixed
with the H-rich circumstellar matter.

\section{Time-dependent emission lines from thermal plasma}

The thermal emission model assumes collisional ionization of atoms and
is applied to rarefied plasmas heated by an external source (the
solar corona provides an example). Photons originating in atomic
transitions, in radiative recombination, etc. leave the
plasma freely. The plasma cooling time depends on density and chemical
composition; for example, for a fully ionized plasma composed of
ions with the charge $Z$,
\begin{equation}
t_c\approx 2\times 10^{15}\, [s]
\left(\frac{T}{10^8}\right)^{1/2}n_i^{-1}Z^{-2}\approx
2\times 10^{15}\, [s]
\left(\frac{T}{10^8}\right)^{1/2}n_e^{-1}Z^{-1}
\label{tcool}
\end{equation}
The numerical factor in this formula is valid with the 20\%-accuracy
for a purely hydrogen plasma. Clearly, this formula gives
an {\it upper limit} for the cooling time in real plasmas.

For densities $n_e \sim 10^{15}\t{cm}^{-3}$, advocated by Reeves et
al., the plasma cools faster than in 1 second and a stationary thermal
model is appropriate. At lower densities, for example,
$n_e\sim10^{11}-10^{12}\t{cm}^{-3}$, the cooling time becomes much
longer than the heating time $t_h\sim 90$ s (the GRB 011211
duration in its rest frame) and time-dependent effects must be
taken into account. The relevant parameter here is $nt$. For large
$nt$, plasma relaxes to a stationary state.

In the case of GRB X-ray afterglows, the non-stationarity has two
faces. On one hand, the real X-ray spectrum could change over the
integration time in the detector ($\sim 5000$ s for GRB 012111)
due to temperature decrease even if the plasma density is high and
the parameter $nt$ formally allows for stationary ionization. On the
other hand, when $nt$ is small, relaxation effects could 
cause ions to appear which were absent in the stationary case. We
stress here that there is strong dependence of plasma cooling time
on the metallicity. The real cooling time depends on ionization
states and abundances of different ions so time-dependent effects
play significant role in low-density plasma.

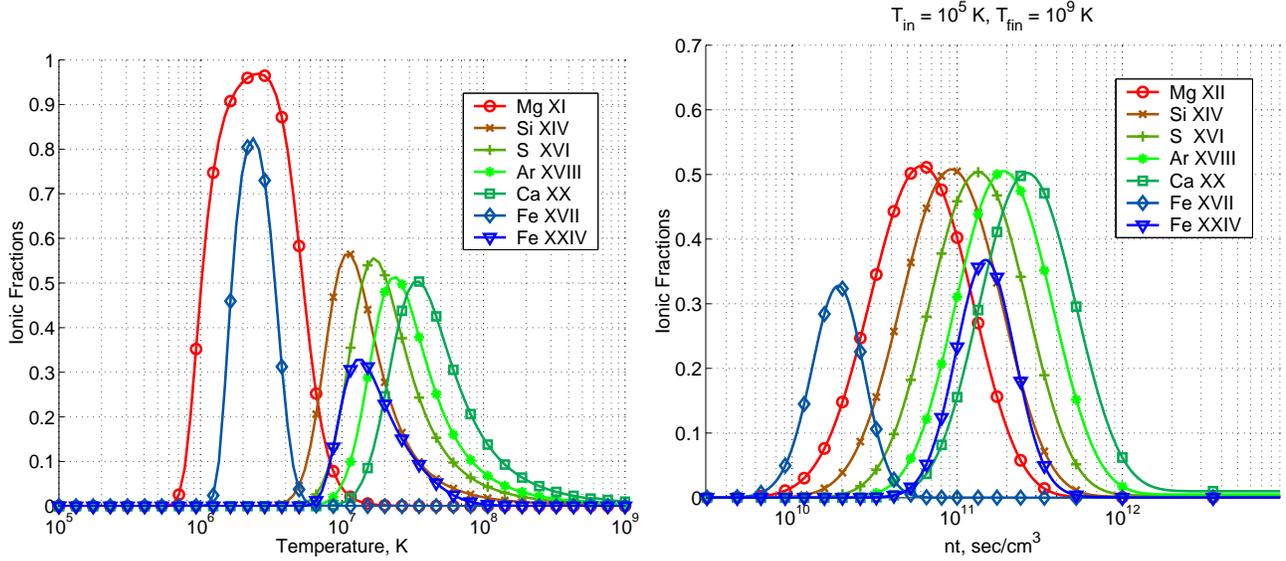
\FIGG abund abunds ntT59 [width=.5\textwidth] Relative number
densities of ions as a function of temperature for stationary case
(left panel) and as a function of time for time-dependent model
(right panel).

This latter face of non-stationarity is illustrated in \rfig{abund}
which shows relative number densities of some ions as a function of
temperature (for stationary case) and as a function of the parameter
$nt$ for time-dependent ionization. In the latter case, an initially
cold plasma ($10^5$ K) with density $n=10^{12}$ cm$^{-3}$ is
instantly heated up to a temperature of $10^9$~K. This figure
indicates that during thermal relaxation of plasma, ionization
state {\it at a given time} can be close to some stationary
ionization state at different fixed temperature.

\section{X-ray afterglow of GRB 011211}

X-ray afterglow of GRB 011211 was observed by {\it XMM-Newton}
$\sim 11$ h after the GRB (Reeves et al. 2002a,b). The emission
lines faded away over a time interval of $10^4$ s. The volume
emission measure as inferred by Reeves et al. from the average
X-ray luminosity $7\times 10^{45}$ erg/s for plasma with normal
cosmic abundance is  $\mathrm{EM}=n^2_eV=10^{69}\t{cm}^{-3}$. As
we discussed above, such an estimate is only the {\it upper limit},
since real plasma can have higher volume emissivity and smaller
cooling time.

\subsection{Thermal model by Reeves et al.}

According to the model by Reeves et al. (2002a), the observed
X-ray spectrum originates in a dense $n_e=10^{15}\t{cm}^{-3}$
shell of radius  $R=10^{15}\t{cm}$ and thickness $h=10^9\t{cm}$
expanding with a velocity of $0.1$ c from GRB, as the mean
redshift of the identified lines ($z=1.88$) implies (the GRB host
galaxy redshift is $z=2.14$). As such a dense plasma cools down in
1 s, the observed fading time of X-ray lines ($t_{\t{l}}=10^4$ s) is
explained by the light retardation from different parts of
instantly illuminated shell (this geometrical effect was invoked
by Lazzati et al. 1999 to explain the X-ray afterglow evolution of
other GRBs).

This model has obvious caveats which make
its applying to GRB 011211 questionable.

\begin{itemize}
\item
{\bf Short plasma cooling time:}\\
If cooling time is as short as 1 s, an overwhelming part of the observed X-ray emission
must be generated during illumination of the plasma by gamma-ray
radiation from the GRB ($t_h=90$ s in the GRB rest frame). The time
evolution is then entirely due to the retardation, i.e. to the geometrical effect.
This implies
that the observed X-ray photons were actually generated {\it
simultaneously} with the powerful
illumination by hard photons fr-om the GRB. Then, in order to explain
the corresponding X-ray spectrum the photoionization (nebular)  approximation but
not the thermal (coronal) one should be used. However, the
photoionization model is claimed to be inadequate (Reeves et al.
2002a). In order to try to explain these observations within the
frame of a thermal plasma model, we must assume much longer cooling time $t_c>t_h$,
hence lower plasma densities.

\item
{\bf Large emission measure:}\\
The EM in this model is determined from the mean X-ray continuum
luminosity and with the assumed density $n_e=10^{15}$ cm$^{-3}$
relates to the layer volume $V=R^2h= (10^{15})^2\times
10^{9}=10^{39}$ cm$^3$ which we observe in 1 s. As the X-ray
luminosity is registered over $10^4$ s, there must be $10^4$ such
volumes! This contradicts the assumption that one and the same
shell of radius $R=10^{15}$ cm and thickness $h=10^{9}$ cm is emmitting. So in
order to save the model, on has to assume the existence of clumps
with size $l=10^{12}\t{cm}\gg h$, which are optically thick at
such densities.

\item
{\bf Apparent absence of iron lines:} \\
The explanation was that iron \fefsx\ does not have enough time to
be produced from the radioactive \nifsx\  expelled by SN. Under
special assumptions about SN~II mechanism one can obtain the deficit of
iron isotopes (Rausher et al. 2001). However, if the plasma shell
in this model is indeed expelled by SN, such an SN is hardly to be
typical -- in particular, it can be asymmetrical. So the initial
amounts of iron, nickel, and cobalt can be comparable. Asymmetric
explosions of SN that may be connected to GRBs give different
results (Pruet et al. 2002). Some model calculations of hypernovae
by Umeda et al. (2002) yield up to 40\% of prompt iron isotopes. So the
iron yields in SN ejecta are strongly model-dependent, and
the apparent absence of iron in the X-ray spectrum can not provide
a reliable argument in favor of a supernova associated with the GRB.

\item
{\bf High density and high velocity of the shell:}\\
Physical characteristics of the SN ejecta as a dense rapidly expanding
shell also meet difficulties. Similar dense shells are produced in
radiation hydrodynamics calculations of SN (Falk and Arnett 1977;
Chevalier and Klein 1979; Blinnikov and Bartunov 1993),
but their properties are quite different. The total energy of
the shell in the model by Reeves et al. approaches
the typical SN explosion energies, while the dense
shells arising in hydro calculations either have smaller masses
or lower velocities, of order of several 1000 km/s (Chugai et al. 2002)
and carry only a small fraction of the SN explosion energy.
\end{itemize}

\subsection{Thermal model with time-dependent ionization}
In our model, the observed X-ray fading time $t_{\t{l}}$ is identified
with thermal relaxation time of a plasma heated up by GRB.

\subsubsection{Emitting region geometry}
Let the GRB occur inside (or nearby) a star forming region of
complex clumpy structure. This assumption is quite natural and
has solid observational support (van Paradijs et al. 2000). In
order for plasma cooling time (\ref{tcool}) to be $10^4$ s, the
gas density must be $n_e\sim 10^{11}-10^{12}\t{cm}^{-3}$. To
obtain the observed EM ($n^2_e\,V \lesssim 10^{69}\t{cm}^{-3}$,
the volume of the emitting region must be $V \lesssim 10^{47} -
10^{45}\t{cm}^3$. At the same time, it must be optically thin for
X- and gamma-ray photons, $nl\sigma_T\sim 1$, where $\sigma_T\sim
10^{-24}\t{cm}^2$ is Thompson electron cross-section. It is hard
to meet both conditions in a homogeneous medium, but a more
realistic rarefied plasma with clumpy structures can do the job.

Indeed, let us consider $\sim 10^6$ gas clouds with density
$\sim 10^{12}\t{cm}^{-3}$ each and a size $\sim 10^{13}\t{cm}$.
The total volume of the dense component in such clouds amounts to
the required value $10^{45}$ cm$^3$ and gives the necessary EM
$=10^{69}\t{cm}^{-3}$ (we recall that these are upper limits; for
example, using realistic plasma cooling function with normal
cosmic abundances yields $\mathrm{EM} \sim 4\times 10^{68}
\t{cm}^{-3}$ and solar plasma without H and He yields $\mathrm{EM}
\sim 10^{65} \t{cm}^{-3}$, see also \rfig{cool}). The total mass
of the emitting clumps is less than $\sim 0.3 M_\odot$. Clumps of
such densities and size are indeed observed as maser condensations
in stellar winds from young stars and late-type stars (e.g. Bains et
al. 2002, Richards et al. 1999, Lekht et al. 2001). The total
volume occupied by this inhomogeneous medium is determined by
energy balance and the width of a gamma-ray beam.

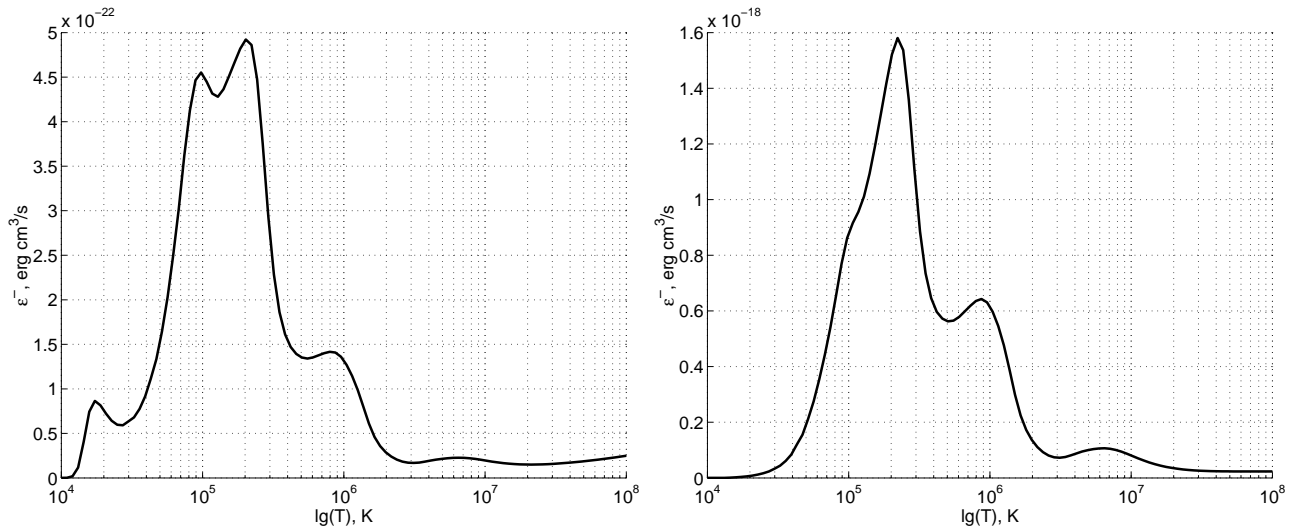
\FIGG cool eloses elosesme [width=.5\textwidth] Plasma cooling
function $\varepsilon^{-}$ (erg cm$^{3}$ s$^{-1}$). {\bf Left
panel}: solar composition, then the emission measure for $T=10^8$ K is $\mathrm{EM} =
n^2\,V = L_X/\varepsilon^{-}\simeq 3.7\times10^{68}\t{cm}^3$; {\bf
Right panel}: metal-rich medium (without H and He), $\mathrm{EM} =
n^2\,V = L_X/\varepsilon^{-} \simeq 1.4\times10^{65}\t{cm}^3$

Heating matter up to $10^8$ K would require energy deposit
$\epsilon_0 = 10\t{keV}$ per nucleon, i.e. the energy
$E_{\t{h}}=10^{12}\t{cm}^{-3}\times
10^{45}\t{cm}^3\times10\,\t{keV} = 10^{48}$ erg is necessary. The
total isotropic energy of GRB 011211 is $E_{\gamma} =
5\times10^{52}$ ergs. The effective cross-section of gamma-ray energy
deposition per nucleon with good accuracy is
$\sigma_\gamma \simeq 0.1\sigma_T$ (Ambwani and Sutherland 1988).
Considering the fluence from the source at a distance $d$ to be
$F=E_{\gamma}/4\pi d^2$, the heating energy per nucleon is
$\epsilon_0=\sigma_\gamma F$, so
$$
  d=\sqrt{\frac{\sigma_\gamma E_{\gamma}}{4\pi\epsilon_0}}\simeq 2\times
  10^{17}\; \t{cm}\; ,
$$
i.e. the GRB energy is sufficient to heat up the surrounding matter up
to distances $d\simeq 0.1$ pc. Note that $F$ plays more important
role than $E_{\gamma}$; it is measured by our detectors (at larger
distances of course), and no assumptions on gamma-ray emission
beaming should be made to estimate $d$.

The gamma-ray beam opening angle
$\theta$ determines the total volume of the illuminated matter.
The simple geometry leads to the estimate
$$
d(1-\cos\theta)=ct_{\t{l}}/(1+z)\simeq 10^{14} \quad \hbox{cm}
$$
Assuming not very large angle (which appears natural in view of
the large value of isotropic $E_{\gamma}$) we have
$d(1-cos\theta)\simeq d\theta^2/2$, i.e. the illuminating gamma-ray
beam size is $R \simeq \theta d = (2d\, ct_{\t{l}}/(1+z))^{1/2} =
(2 \times 10^{14}\t{cm} \times d)^{1/2}$. For example, $R\simeq
10^{16}$ cm for $d=2\times 10^{17}$ cm and $\theta \simeq R/d =
0.05$, i.e. $3^\circ$  (the total beam angle is twice as large,
$2\theta=6^\circ$ ).

The total volume of the medium illuminated by GRB is quite large,
$V \sim R^2 d = 3 \times 10^{49}$ cm$^{-3}$. We stress here that
though gamma-ray photons from GRB fly this distance $d$ in several
months, they reach the detector simultaneously with X-ray photons
they created passing the clouds, so X-ray emission comes not from
a dense shell $10^{14}$ cm in thickness but from an extended
clumpy region with the size of order of $d$. The filling factor of
this medium can be $10^{-4}$ or larger if the condensations are located
in a relatively thin layer and not across the entire distance $d$.
This simple estimate shows that the physically acceptable structure of
a clumpy ISM can actually be the source of the observed X-ray
emission.

The necessity of clumpy structure of emitting plasma was also
obtained by Lazzati (2002) from independent analysis of stationary
thermal X-ray afterglows from GRBs. We stress again here that in
contrast to Lazzati, we do not require the shell around the GRB,
for us it suffices that the medium with the required physical
parameters lies in the cone of the gamma-ray beam. This plasma can be
not genetically connected with GRB. In our model, orphan X-ray
afterglows can well show transient emission features like those
observed in the case of GRB 011211.

\subsubsection{Spectral modelling}
To model X-ray spectra from thermal plasma, including the case of
GRB 011211, we employ a numerical code originally worked out by P.
Lundqvist for calculations of collisional ionization of a
stationary plasma, which takes into account the following basic
processes: ionization by electron impact, autoionization,
photorecombination, dielectronic recombination, and charge
transfer. The algorithm  outputs bremsstrahlung (free-free),
free-bound, bound-bound, and two-photon emission. The code has
been upgraded by S.I.~Blinnikov and E.I.~Sorokina to a
time-dependent version ($\dot n_i = f(T(t),n(t),Z,...)$ for all
ionic species). Input parameters are initial temperature, density,
and abundances of elements.

To calculate time-dependent spectra, the temperature dependence on time
$T(t)$ is needed. The plasma cooling function $\varepsilon^{-}$ (erg
cm$^{3}$ s$^{-1}$) is almost independent of temperature for
$2\times 10^6$K -- $10^8$K  (see  \rfig{cool}). Then the
temperature change with time can be found from the differential
equation
$$
\frac{dT}{dt} = -\frac{\varepsilon^{-}(T_0)n_e^2}{C_V\rho}
$$
where $C_V=(3/2){\cal R}(1/A+X_e)$ is heat capacity per unit
mass at fixed volume, ${\cal R}$ is the universal gas constant,
$A$ is the mean mass of ions in atomic units, $X_e$ is the number
of free electrons per nucleon, $T_0 = 10^8$K is the initial
temperature. Then in the above temperature range the temperature
decreases linearly with time
\begin{equation}
\label{T(t)}
T(t) = T_0-\varepsilon^{-}(T_0)n_e^2\,t/(C_V\rho)
\end{equation}

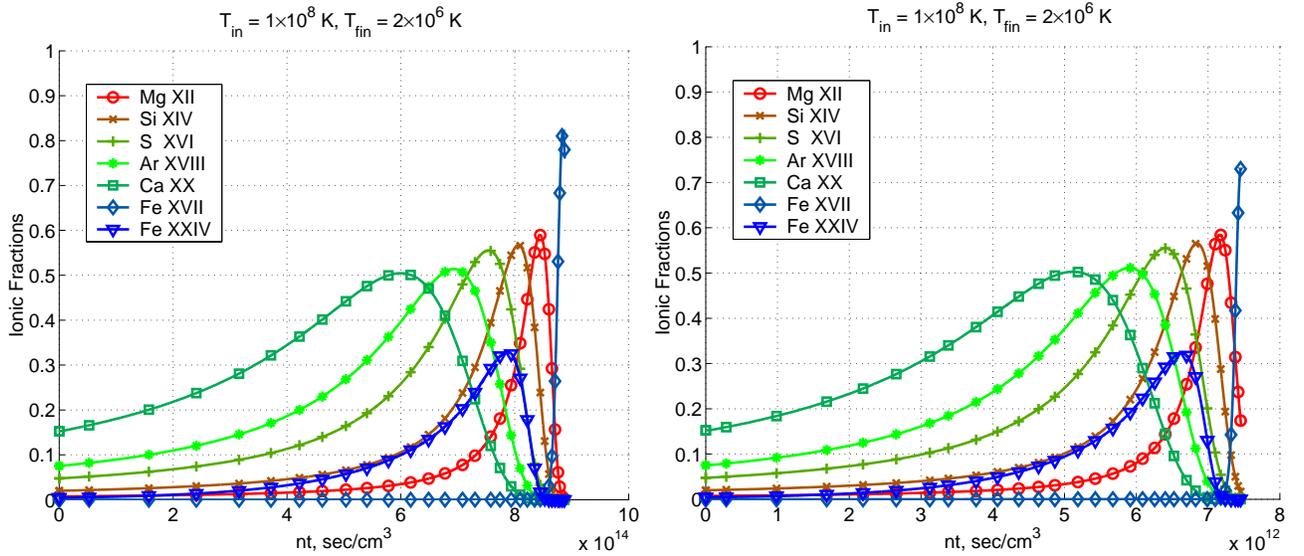
\FIGG nt ntT86 ntT86me [width=.5\textwidth] Relative number
densities of ions as a function of the parameter $nt$ for linear
plasma cooling law (\protect\ref{T(t)}). {\bf Left panel:} solar
composition; {\bf right panel:} metal-rich plasma (without H and
He)

The time evolution of ion number densities (\rfig{nt}) is
calculated numerically and determines the appearance of the
corresponding emission X-ray lines (\rfig{sp}). These figures show
strong time evolution of thermal spectra from optically thin
cooling plasma initially heated by GRB up to 10-keV temperatures.
For normal cosmic abundances, emission lines from moderately
ionized iron atoms immediately appear. As plasma cools down, lines
of H-like and He-like iron emerge (\rfig{sp}). If metal enriched
(and, correspondingly, with deficit of H and He) SN ejecta is
observed, emission lines from metals become more prominent against
the continuum. At late cooling stages, photorecombination edges of
metals prevail (\rfig{sp}).

\FIGG sp {spz214nh1d1e+012} {spzme214nh1d1e+010}
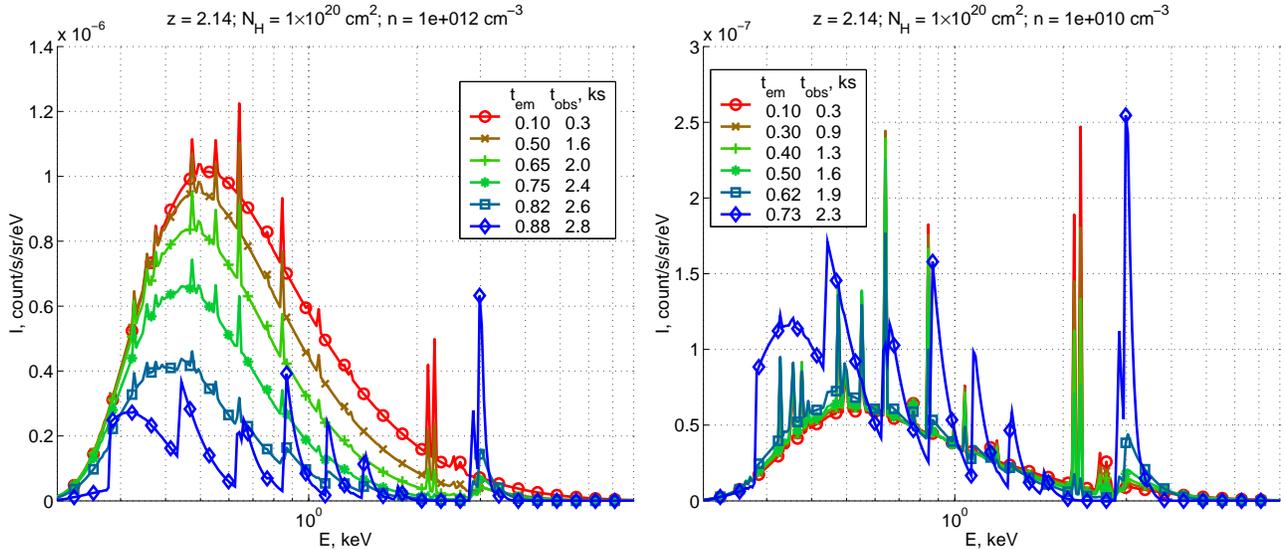
[width=.5\textwidth] Time evolution of X-ray spectrum (in the
observer frame at the GRB redshift $z=2.14$) for different time
moments. {\bf Left panel:} solar composition; {\bf right panel:}
metal-rich plasma (without H and He)

\section{Discussion}
Analysis of our calculations of emission spectra from thermal
optically thin plasma suggests that interpretation of early X-ray
GRB afterglows in term of thermal model requires (1) a moderate
density of the emitting plasma ($n_e\sim
10^{11}-10^{12}\t{cm}^{-3}$), and  (2) a clumpy structure of the
circumstellar (interstellar) medium. In our model, X-ray line
emission forms not necessarily in matter genetically related to
gamma-ray burst source, it is sufficient for it to lie within the
gamma-ray illumination beam at distances $\lesssim 0.1$ pc from
the GRB site. Specifically, for the observed parameters of GRB
011211, $L_X(0.6-30)\,\hbox{keV}\approx 7\times 10^{45}$ erg/s (in
the source rest-frame), a total of $\sim 0.3 M_\odot$ of matter
comprised in clouds $\sim 1$ AU across inside a volume of $\sim
10^{49}\t{cm}^3$ illuminated by gamma-ray beam with the full width
$2\theta\sim 6^\circ$ would suffice. Note that such narrow angles
are generally obtained in the standard model interpretation of
time decay of GRB afterglows (Frail et al. 2001). The parameters
of individual clouds are also astrophysically feasible and
correspond to dense maser condensations in stellar winds from AGB
stars (Bains et al. 2002, Richards et al. 1999, Lekht et al. 2001).

The proposed model interprets the observed temporal evolution of
X-ray spectrum in terms of thermal relaxation time of ions and
plasma cooling, and does not appeal to purely geometrical
arguments. However, the comparison of the calculated spectra with
those observed by Reeves et al. does require enhanced
$\alpha$-metal abundances. The apparent absence of iron lines
either requires its underabundance, which is less probable
according to current SN nucleosynthesis, or can be explained by
their rapid variability or washing out against a high level of
nonthermal power-law continuum (which is not excluded by spectral
fitting).

Thermal X-ray spectra of a
metal-rich shell ejected by an SN preceding GRB for a few days
would rather show photo-recombination edges and not stationary
emission lines with time evolution due to geometrical factor.

In our model, it appears more natural to consider that no
blue-shifted X-ray lines were observed, if the clumpy medium is a
circumstellar envelope produced by slow superwind from the GRB
progenitor. The case is even stronger if what we observe is a
nearby clumpy cloud in a star-forming region. If the significant
blue-shift of the X-ray lines will be confirmed, it will provide a
strong evidence that the GRB is preceded by a supernova. However,
if these rapidly moving clouds are hydrogen rich, they can hardly
be located at such small distances as inferred by Reeves et al.
(2002a) and the preceding supernova must occur several months or
even years before the GRB event (the supranova scenario, Vietri
and Stella 1998).

Apparently, the adequate explanation of the observed spectrum of
X-ray afterglow of GRB 011211 should be sought for within the
frames of a combined model of photoionization heating of an
inhomogeneous medium around the GRB progenitor and time-dependent
collisional ionization in more dense condensations with an optical
thickness for scattering about 1. Clearly, the calculation of
realistic X-ray spectra from such media requires appropriate
account for radiative transfer effects, which are beyond the scope
of the present work.

\section{Conclusion}

We have calculated thermal X-ray emission spectra of optically
thin thermal plasma using a time-dependent treatment of collisional
ionization. We show that an instant heating of this plasma up to
the 10-keV temperatures produces time-dependent X-ray emission
spectra in which lines of $\alpha$-element metal ions and iron at
different ionization states are apparent. We propose a model of
thermal X-ray emission for the observed early X-ray afterglow of
GRB 011211, in which a narrow gamma-ray beam passes through clumpy
interstellar medium, not obligatorily genetically related to GRB.
The non-stationarity of the X-ray spectrum is related not to
a geometrical factor (time retardation from different parts of the
shell), but to a physical non-stationarity of the collisional
ionization and thermal relaxation of plasma with density of order
$10^{11}-10^{12}$ cm$^{-3}$. Parameters of such a medium resemble
maser condensations in active star-forming regions. We conclude
that the account for time-dependent effects in plasma is
unavoidable in spectral X-ray diagnostics of the medium around GRB
sites within the frame of thermal plasma models.

The work is supported by RFBR grant 02-02-16500. D.I.~Kosenko
received a partial support through RFBR grants 02-02-27445z and
01-15-99310m. S.I.~Blinnikov and E.I.~Sorokina are partially
supported by RFBR grant 00-02-17230 and thank W.~Hillebrandt
and other colleagues at MPA (Garching, Germany) for their hospitality
and support. They and P.~Lundqvist also acknowledge support from Wenner-Gren Science
Foundation and Royal Academy of Sweden.

\end{document}